\documentclass[twocolumn]{webofc}
\usepackage[varg]{txfonts}   
\usepackage[utf8]{inputenc}
\usepackage{siunitx}

\newcommand{\fg}[1]{Fig.\,\ref{fig:#1}}
\newcommand{\eq}[1]{Eq.\,\ref{eq:#1}}
\newcommand{\lna}{\ln\!A}
\newcommand{\mlna}{\langle \lna \rangle}
\newcommand{\xmax}{X_\text{max}}
\newcommand{\nmu}{{N_\mu}}

\begin{document}
\title{Report on Tests and Measurements of Hadronic Interaction Properties with Air Showers}

\author{\firstname{H.P.} \lastname{Dembinski}\inst{1}\fnsep\thanks{\email{hdembins@mpi-hd.mpg.de}} \and
        \firstname{J.C.} \lastname{Arteaga-Vel\'azquez}\inst{2} \and
        \firstname{L.} \lastname{Cazon}\inst{3} \and
        \firstname{R.} \lastname{Conceição}\inst{3} \and
        \firstname{J.} \lastname{Gonzalez}\inst{4} \and
        \firstname{Y.} \lastname{Itow}\inst{5} \and
        \firstname{D.} \lastname{Ivanov}\inst{6} \and
        \firstname{N.N.} \lastname{Kalmykov}\inst{7} \and
        \firstname{I.} \lastname{Karpikov}\inst{8} \and
        \firstname{S.} \lastname{M\"{u}ller}\inst{9} \and
        \firstname{T.} \lastname{Pierog}\inst{9} \and
        \firstname{F.} \lastname{Riehn}\inst{3} \and
        \firstname{M.} \lastname{Roth}\inst{9} \and
        \firstname{T.} \lastname{Sako}\inst{10} \and
        \firstname{D.} \lastname{Soldin}\inst{4} \and
        \firstname{R.} \lastname{Takeishi}\inst{11} \and
        \firstname{G.} \lastname{Thompson}\inst{6} \and
        \firstname{S.} \lastname{Troitsky}\inst{8} \and
        \firstname{I.} \lastname{Yashin}\inst{12} \and
        \firstname{E.} \lastname{Zadeba}\inst{12} \and
        \firstname{Y.} \lastname{Zhezher}\inst{7}
        for the EAS-MSU\inst{13}, IceCube\inst{14}, KASCADE-Grande\inst{15}, NEVOD-DECOR\inst{16}, Pierre Auger\inst{17}, SUGAR\inst{18}, Telescope Array\inst{19}, and Yakutsk EAS Array\inst{20}\ collaborations
}

\institute{%
Max Planck Institute for Nuclear Physics, 69117 Heidelberg, Germany
\and
Institute of Physics and Mathematics, Universidad Michoacana, C.P. 58040 Morelia, Michoacan, Mexico
\and
LIP, Lisbon, Portugal
\and
Bartol Institute, University of Delaware, Delaware, USA
\and
Nagoya University, Nagoya, Japan
\and
High Energy Astrophysics Institute and Department of Physics and Astronomy, University of Utah, Salt Lake City, Utah, USA
\and
D.V. Skobeltsyn Institute of Nuclear Physics, M.V. Lomonosov Moscow State University, Moscow 119991, Russia
\and
Institute for Nuclear Research of the Russian Academy of Sciences, 60th October Anniversary Prospect 7a, Moscow 117312, Russia
\and
Karlsruhe Institute of Technology, Eggenstein-Leopoldshafen, Germany
\and
Institute for Cosmic Ray Research, University of Tokyo, Tokyo, Japan
\and
Department of Physics, Sungkyunkwan  University, Jang-an-gu, Suwon, Korea
\and
National Research Nuclear University MEPhI (Moscow Engineering Physics Institute), Moscow 115409, Russia
\and
EAS-MSU experiment, Moscow, Russia
\and
IceCube Neutrino Observatory, Madison WI, USA \\
\emph{Full author list:} \url{https://icecube.wisc.edu/collaboration/authors/icecube}
\and
KASCADE-Grande experiment, Karlsruhe, Germany \\
\emph{Full author list:} \url{https://web.ikp.kit.edu/KASCADE}
\and
NEVOD-DECOR experiment, Moscow, Russia
\and
Pierre Auger Observatory, Malargüe, Argentina \\
\emph{Full author list:} \url{http://www.auger.org/archive/authors_2018_10.html}
\and
SUGAR Array, Sidney, Australia
\and
Telescope Array Project, Salt Lake City UT, USA \\
\emph{Full author list:} \url{http://www.telescopearray.org/research/collaborators}
\and
Yakutsk EAS Array, Yakutsk, Russia \\
\emph{Full author list:} \url{https://ikfia.ysn.ru/en/theyakutskarrayteam}
}

\abstract{%
We present a summary of recent tests and measurements of hadronic interaction properties with air showers. This report has a special focus on muon density measurements. Several experiments reported deviations between simulated and recorded muon densities in extensive air showers, while others reported no discrepancies. We combine data from eight leading air shower experiments to cover shower energies from PeV to tens of EeV. Data are combined using the $z$-scale, a unified reference scale based on simulated air showers. Energy-scales of experiments are cross-calibrated. Above 10 PeV, we find a muon deficit in simulated air showers for each of the six considered hadronic interaction models. The deficit is increasing with shower energy. For the models EPOS-LHC and QGSJet-II.04, the slope is found significant at 8 sigma.}
\maketitle
\section{Introduction}
\label{intro}

\begin{figure}
\includegraphics{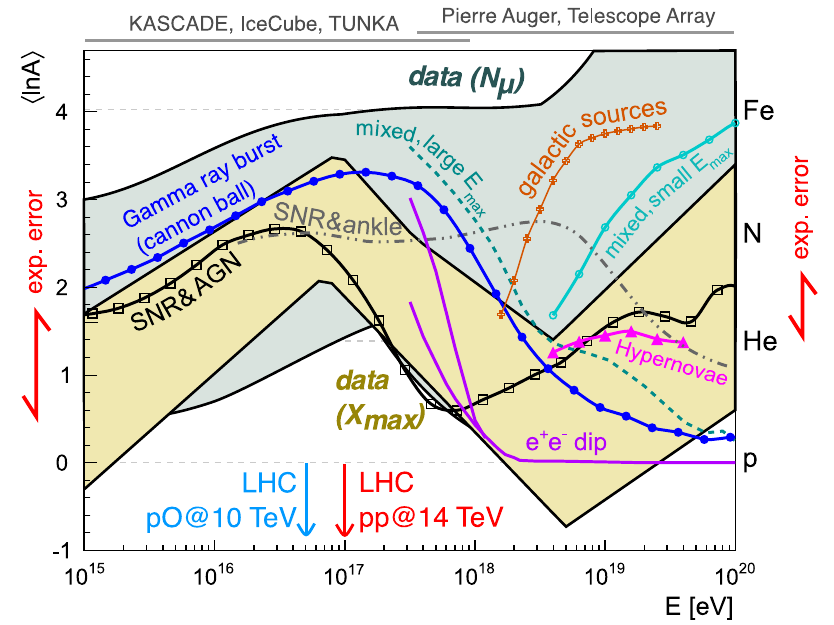}
\caption{Mass composition of cosmic rays quantified by $\mlna$ as a function of cosmic-ray energy $E$. Model predictions (markers and lines) are compared to data bands, all taken from the review by Kampert and Unger~\cite{Kampert:2012mx}. Vertical arrows at the sides indicate the instrumental error achieved by leading experiments at low and high energies. The figure is discussed in the text.}
\label{fig:kampert_unger}
\end{figure}

Cosmic rays with energies larger than $10^{15}\,\text{eV}$ can only be indirectly observed via extensive air showers. A detailed understanding of the hadronic physics in an air shower is needed to infer the energy and mass of the cosmic ray from air shower measurements. The fluorescence technique has reduced the model-dependence for the energy measurement by tracking the longitudinal shower development, but inferring the mass accurately is still a challenge.

The energy-dependent mass composition of cosmic rays carries a unique imprint from the origin and propagation of cosmic rays. Attempts to measure the mass composition are complementary to direct searches for cosmic-ray sources via coincident observation in a multi-messenger approach or statistical correlation with potential sources. In Fig.~\ref{fig:kampert_unger}, predictions (lines and markers) are shown for the mean-logarithmic-mass $\mlna$ from different theories. A measurement to an accuracy of 10\,\% of the proton-iron difference is technically possible, but uncertainties in hadronic interactions prevent the full exploitation of existing measurements. The two leading observables to infer $\mlna$ are the depth $\xmax$ of the shower maximum in the atmosphere (yellow band in \fg{kampert_unger}), and the number $\nmu$ of muons produced in the shower (green band in \fg{kampert_unger}). The width of those bands has two main contributions: the experimental uncertainties, and the uncertainties inherent in converting the air shower observables into $\mlna$, which requires air shower simulations with hadronic interaction models. Leading models are EPOS-LHC~\cite{Pierog:2013ria}, QGSJet-II.04~\cite{Ostapchenko:2013pia}, and SIBYLL-2.3c~\cite{Riehn:2017mfm}. These models are called post-LHC models, due to their tuning to LHC data. Their variation is dominating the uncertainty. Moreover, the mass composition obtained with $\xmax$ and $\nmu$ is not consistent, which implies that the models are not correctly describing all aspects of hadronic physics in air showers.

Tests and measurements of hadronic interaction properties with air showers address these issues. They can reduce uncertainties in modeling the air shower development, so that accurate estimates of the cosmic-ray mass can be obtained. They also offer opportunities to test the standard model of particle physics under extreme conditions. The cms-energy in the first interaction of an air shower initiated by a $10^{20}$\,\si{eV} cosmic ray is \SI{432}{TeV} in the nucleon-nucleon system, 33 times higher than what is currently accessibly at the LHC.

In this report, we will review recent tests and measurements of hadronic interaction properties, starting with the measurements of the electromagnetic shower component, but then focusing on muon density measurements. There is a long-standing problem with the correct simulation of muons in air showers. The HiRes/MIA collaboration already reported a discrepancy in simulated and measured air showers between $10^{17}$ to $10^{18}$\,eV in the year 2000~\cite{AbuZayyad:1999xa}. The NEVOD-DECOR experiment reported an increase of muon density relative to simulation from $10^{15}$ to $10^{18}$\,eV in 2010~\cite{Bogdanov:2010zz}. Above $10^{17}$\,eV, an excess of multi-muon events at energies around $10^{18}$\,eV over the expectation was observed, and a muon deficit in simulations was suggested as an explanation. The experiments KASCADE-Grande~\cite{Apel:2017thr} and EAS-MSU~\cite{Fomin:2016kul} reported no muon discrepancy in this energy range when the latest hadronic interaction models tuned to LHC-data are used to simulate air showers, while the SUGAR array~\cite{Bellido:2018toz} reported a muon deficit even for these models. The Pierre Auger Observatory~\cite{Aab:2014pza,Aab:2016hkv} and Telescope Array~\cite{Abbasi:2018fkz} also observed a muon deficit in $10^{19}$\,\si{eV} showers simulated with the latest models. These measurements, preliminary data from IceCube~\cite{Gonzales:2018IceTop} and AMIGA~\cite{mueller2018}, and unpublished data from Yaktusk~\cite{yakutskpc} are systematically compared in this report.


\section{Measurements of electromagnetic shower component}

Most measurements of the electromagnetic component of air showers show good or acceptable agreement with simulations, especially when the post-LHC generation of hadronic interaction models is used. We list recent measurements briefly and focus on deviations.

\begin{description}
\item [Proton-air cross-section] The proton-air cross-section has been measured based on the slope of the tail of the $\xmax$-distribution~\cite{Ulrich:2015yoo,Abbasi:2015fdr} by the Pierre Auger Observatory and Telescope Array. Conceptually, this is the most direct measurement of a hadronic interaction property with air showers, the dependence of the analysis on uncertainties in the mass-composition and the shower development is reduced. The measurements start to constrain hadronic interaction models.

\item [Moments of $\xmax$-distribution] The first two moments of the $\xmax$ distribution measured by the Pierre Auger Observatory have been mapped to the first two moments of the $\lna$-distribution with parameters from simulations~\cite{Bellido:2017cgf}. Some unphysical second moments were found for QGSJet-II.04.

\item [Longitudinal shape] The average longitudinal shapes of air showers recorded by the Pierre Auger Observatory have been compared to simulations, more specifically the width of the profile around the maximum and the asymmetry of the rising and falling edge~\cite{Aab:2018jpg,UHECRemProfiles}. The measurements are compatible with simulations using post-LHC models. 

\item [Lateral density profile] The slope of the lateral density profile of electrons and photons is sensitive to the cosmic-ray mass. At the IceCube Neutrino Observatory, measurements of the slope with the surface detector array in 2.8\,km altitude a.s.l.\ were compared to measurements of muon bundles below a kilometer of ice, which are also sensitive to the mass~\cite{DeRidder:2017alk}. SIBYLL-2.1~\cite{Ahn:2009wx} and EPOS-LHC are inconsistent with the combined measurements.

\item [Attenuation with zenith angle] The attenuation of the signals measured at ground with increasing zenith angle have been compared by Telescope Array up to 45$^\circ$~\cite{ivanov2018}, and up to 40$^\circ$ by KASCADE-Grande~\cite{Apel:2017thr}. No deviations to simulations were found.
\end{description}

\section{Measurements of muonic shower component}

Most measurements of the muonic component of air showers show disagreement to simulations. The deviations are difficult to reduce to a single cause, further research on muons in air showers is needed.

\begin{description}
\item [Lateral density] A large body of data on lateral density measurements is available and has been converted into a comparable format. The comparison of lateral density measurements will be discussed in the next section.

\item [Production depth or height] The muon production height has been inferred from signal arrival times in ground detectors of the Pierre Auger Observatory~\cite{Aab:2014dua}, and via a muon tracking detector within KASCADE-Grande~\cite{Apel:2011zz}. The Pierre Auger Observatory converts the measured heights into equivalent slant depths, which is recommended, since hadronic cascades develop along slant depth. Disagreement to simulations are found for QGSJet-II.02~\cite{Ostapchenko:2005nj} at PeV energies, and for QGSJet-II.04 and EPOS-LHC at EeV energies.

\item [Attenuation with zenith angle] The attenuation of the muon lateral profile with growing zenith angle has been measured by KASCADE-Grande up to 40$^\circ$~\cite{Apel:2017thr}. Simulations show larger attenuation factors than data for the models QGSJet-II.02, QGSJet-II.04, SIBYLL-2.1 and EPOS-LHC. The attenuation has a dependence on the lateral distance to the shower axis, which is not reproduced by simulations.

\item [Multiplicity in muon bundles] The ALICE experiment has recently measured the multiplicity in muon bundles~\cite{ALICE:2015wfa}. The relative abundance of high and low multiplicity events is well reproduced by QGSJet-II.04.

\item [Atmospheric flux of TeV muons] The IceCube Neutrino Observatory has measured the atmospheric flux of TeV muons~\cite{Aartsen:2015nss,Fuchs:2017nuo}. The total flux is well reproduced by SIBYLL-2.1. Some tension is found in the zenith-angle dependent flux, which could potentially be fixed by adding charmed mesons and their decays to the simulations. Contributions of charmed mesons are negligible for most air shower measurements, but contribute significantly to multi-TeV muons and neutrinos.

\item [Lateral separation of TeV muons] The IceCube Neutrino Observatory has observed events with laterally separated muons from muon bundles~\cite{Soldin:2018vak}. The events are interpreted as single muons with high transverse momentum, which dominantly originate from the first interaction in an air shower. The statistical distribution of the lateral separation is related to the transverse momentum distribution of hadrons produced in the first interaction. Partial agreement is found for SIBYLL-2.1 and 2.3, disagreement for EPOS-LHC and QGSJet-II.04.

\item [Rise-time of shower front] The Pierre Auger Observatory has published mass-composition measurements based on the normalized rise-time~\cite{Aab:2017cgk} and the rise-time asymmetry~\cite{Aab:2016enk}. The rise-time is the time interval in which the collected surface detector charge rises from 10\,\% to 50\,\% of the final value. Muons in an air shower tend to arrive before electrons and photons, so the rise-time is a tracer of the muon content of the shower. Both analyses show disagreement between data and simulations for QGSJet-II.04 and EPOS-LHC. At lateral distances larger than 1000\,m from the shower axis, however, the rise-time asymmetry is compatible with QGSJet-II.04 predictions.
\end{description}

\section{Measurements of muon lateral density}

\begin{figure*}
\centering
\includegraphics[width=0.325\textwidth,clip,trim=45 10 10 0]{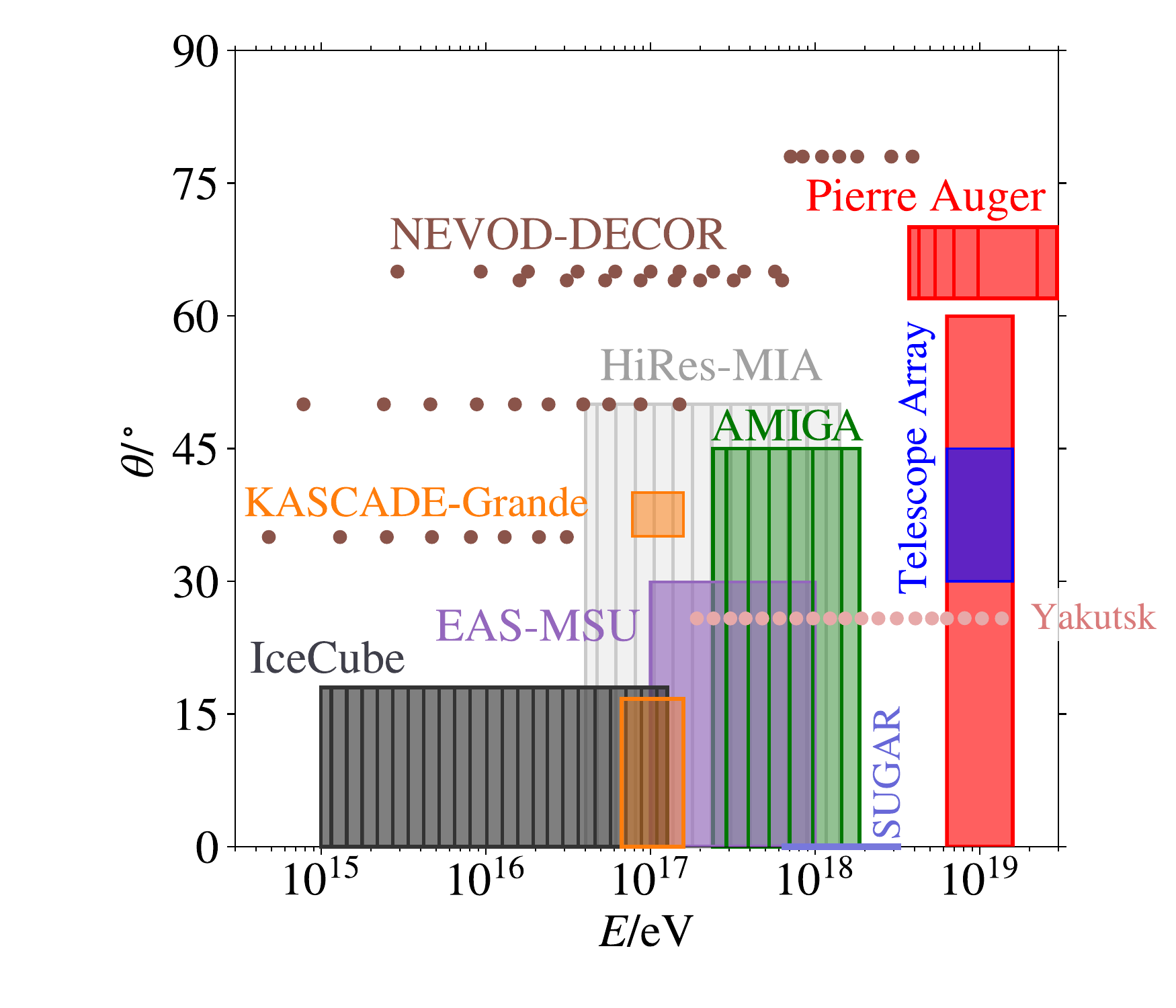}
\hfill
\includegraphics[width=0.31\textwidth,clip,trim=45 10 30 0]{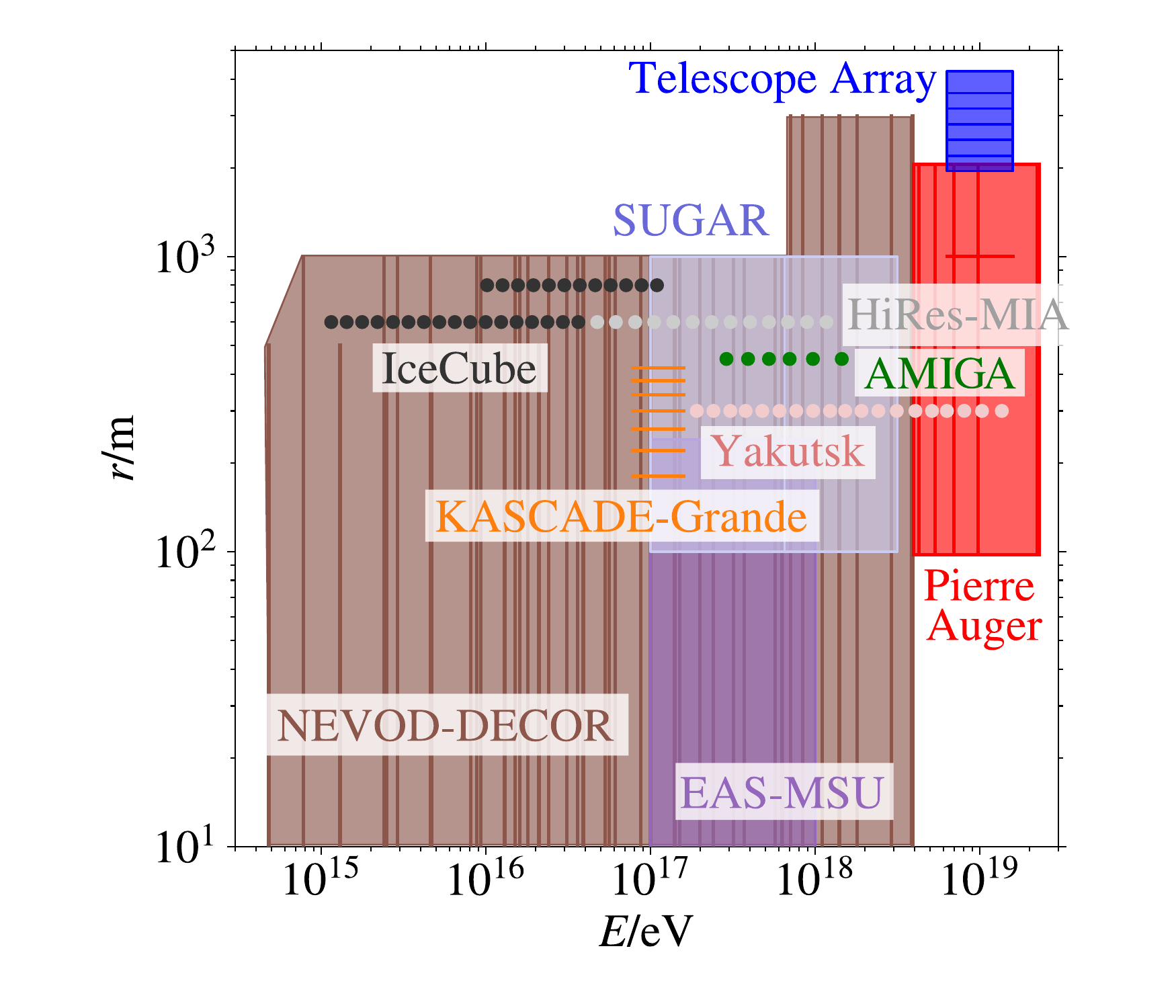}
\hfill
\includegraphics[width=0.32\textwidth,clip,trim=30 10 30 0]{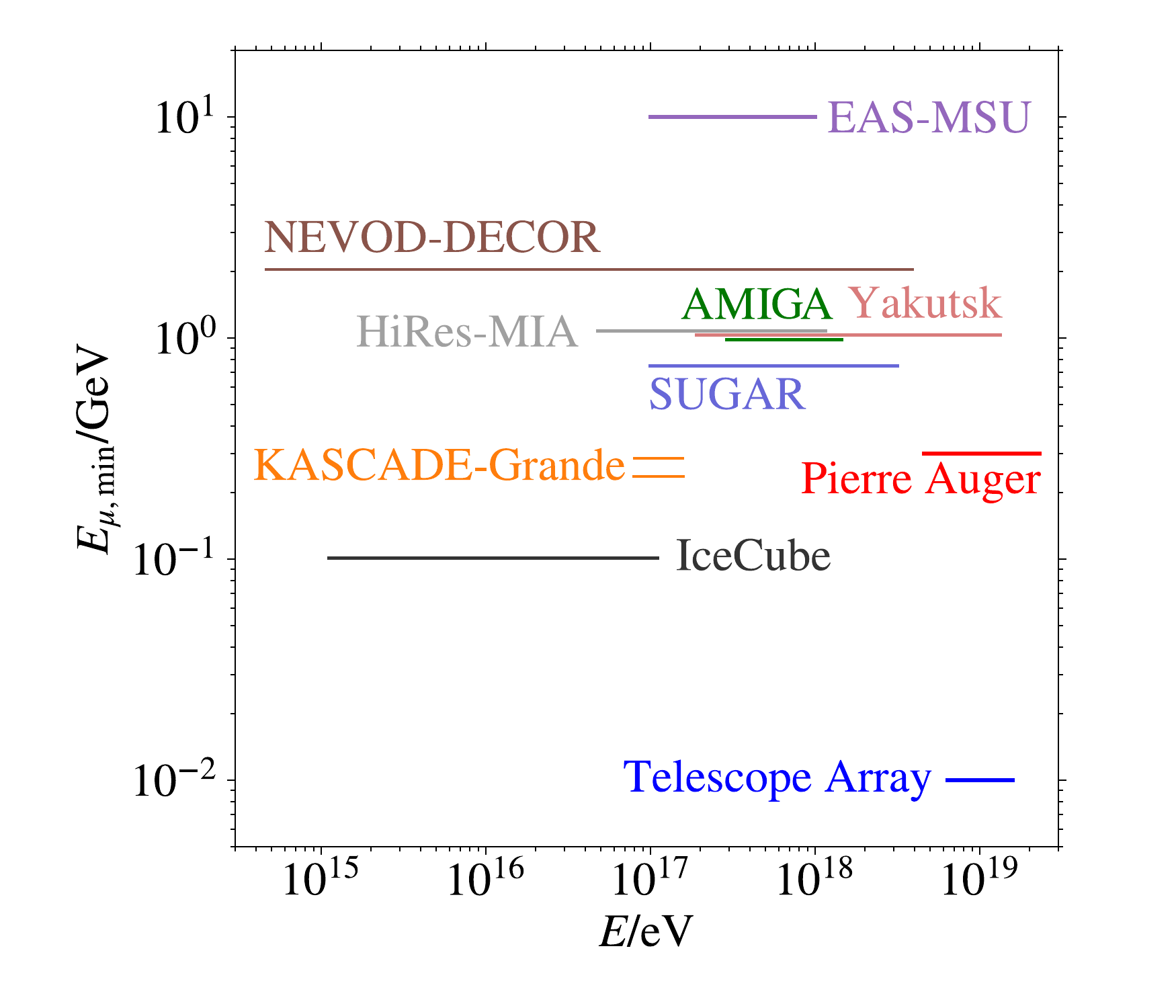}
\caption{Air shower experiments have measured the muon density at ground under various conditions, which are shown here. Points and lines indicate a measurement in a narrow bin of the parameter, while boxes indicate integration over a parameter range. \emph{Left:} Zenith angle of air showers versus shower energy. \emph{Middle:} Lateral distance of the muon density measurement versus shower energy. \emph{Right:} Energy threshold for the muons that are counted in the experiment. Some experiments measure muons below a shielding, which increases the muon energy threshold.}
\label{fig:phase_space}
\end{figure*}

We are analyzing the following measurements of the lateral muon density from eight cosmic-ray experiments:
\begin{itemize}
  \item EAS-MSU~\cite{Fomin:2016kul}
  \item IceCube Neutrino Observatory~\cite{Gonzales:2018IceTop}
  \item KASCADE-Grande~\cite{Apel:2017thr}
  \item NEVOD-DECOR~\cite{Bogdanov:2010zz,Bogdanov:2018sfw}
  \item Pierre Auger Observatory \& AMIGA~\cite{Aab:2014pza,Aab:2016hkv,mueller2018}
  \item SUGAR~\cite{Bellido:2018toz}
  \item Telescope Array~\cite{Abbasi:2018fkz}
  \item Yakutsk, based on preliminary unpublished data~\cite{yakutskpc}.
\end{itemize}
We further show HiRes-MIA data~\cite{AbuZayyad:1999xa} for comparison in several plots, but exclude them from the final results. The HiRes-MIA result systematically differs from all more recent measurements and we did not succeed in contacting one of the authors to better understand the differences.

A direct comparison of muon measurements is not possible, since the muon measurements are performed under very different conditions and using different techniques. The muon density at the ground depends on many parameters which differ from experiment to experiment:
\begin{itemize}
  \item Cosmic-ray energy $E$,
  \item Zenith angle $\theta$,
  \item Shower age (depends on altitude of the experiment, local atmosphere, and zenith angle of the shower),
  \item Lateral distance $r$ from shower axis,
  \item Energy threshold $E_{\mu,\text{min}}$ of the detectors for muons.
\end{itemize}
Since direct comparisons of the measured muon density are unfeasible, each experiment usually compares to air shower simulations. The data/MC ratio is comparable between different analyses and different experiments. In a way, air shower simulations provide a universal reference. The caveat of this approach is that two measurements are only comparable, if simulations with the same hadronic interaction model are available for both.

How the measurements cover the space of parameters is shown in \fg{phase_space}. The measurements as a whole cover most of the parameter space. This is very valuable, since it allows us to detect a possible dependence of the data/MC ratio along all dimensions of the parameter space.

Comparing data/MC ratios instead of just the data introduces a complication. The value of the ratio depends on how the corresponding air shower simulations are selected. While most of the shower parameters can be easily matched in simulation and experiment, the cosmic-ray energy $E$ is difficult to match. This has a large impact. According to the Matthews-Heitler model of air showers~\cite{Matthews:2005sd}, the muon number depends on the energy $E$ and mass $A$ of the cosmic ray in the following way
\begin{equation}
  \nmu = A \left( \frac{E}{A \, C} \right)^\beta = A^{1-\beta}\left(\frac E C \right)^\beta,
  \label{eq:nmu_e_a}
\end{equation}
with power-law index $\beta \simeq 0.9$ and energy constant $C$. The muon number scales almost linearly with the cosmic-ray energy, and with a small power of the mass. For an easier discussion, we take the logarithm on both sides of \eq{nmu_e_a} and compute the mean, which gives a linear equation
\begin{equation}
  \langle \ln\!\nmu \rangle = (1-\beta) \, \mlna + \beta \, \langle \ln(E/C) \rangle.
  \label{eq:lnnmu_e_a}
\end{equation}
Air shower experiments usually have independently calibrated energy scales with systematic uncertainties of 10\,\% to 20\,\%. Two otherwise identical experiments with an energy-scale offset of 20\,\% would find a 18\,\% offset in the data/MC ratios, simply because equivalent measurements are compared to air showers simulated at different (apparent) energies. Cross-calibrating the energy-scales removes these offsets.

When $\langle \ln\!\nmu \rangle$ is directly compared to simulations, the value of $\mlna$ matters. The situation is better when comparing data from different experiments. The value of $\mlna$ is a function of the energy $E$ only, so the effect on two experiments is the same, if both experiments compute the muon density from an unbiased sample of air showers, meaning that cosmic rays are included in the sample with a probability independent of their mass $A$~\cite{Dembinski:2015wqa}. If the sample has a different mass composition $\widehat\mlna$, an offset $(1-\beta) (\widehat\mlna - \mlna)$ is introduced in a comparison to an unbiased sample. A poor detector resolution or using wide energy bins can also introduce biases~\cite{Dembinski:2017kpa}.

Based on this discussion, we can classify the experiments by the measured observables into three groups.

\begin{description}
  \item[Shower energy and muon density] The Pierre Auger Observatory \& AMIGA, Telescope Array, and the Yakutsk experiment are all capable of measuring the shower energy $E_\text{cal}$ using Cherenkov or fluorescence light, which can be converted into the primary energy $E$ with a low model-dependence. These experiments come close to measuring the primary cosmic-ray energy $E$ independently of the muon density at the ground, and can compute the data/MC ratio for showers with the same energy.

  The IceCube Neutrino Observatory also falls into this category, although it does not observe the air shower optically. It can measure the shower energy with low model-dependence thanks to its high altitude, which places the detector close to the average depth of the shower maximum~\cite{Aartsen:2013wda}.

  \item[Muon and electron density] KASCADE-Grande and EAS-MSU measure signals from electrons and muons separately. These experiments compute the data/MC ratio for showers in the same electron-density interval, not for showers in the same cosmic-ray energy interval. The electron density is correlated to the cosmic-ray energy, but also to the muon density~\cite{Kampert:2012mx}. Computing the data/MC ratio for showers in the same electron density interval is conceptually different from the previous case, since \eq{nmu_e_a} and \ref{eq:lnnmu_e_a} do not apply. The data/MC ratios computed by the first and second class are not directly comparable.

  \item[Muon density only] NEVOD-DECOR and SUGAR are pure muon detectors, without a separate energy estimator. The data/MC ratios are again computed differently. The flux of showers is measured in intervals of an event-wise muon density estimate. The measured flux is then compared with a simulated flux in the event-wise muon density estimate, computed from an external model of the cosmic-ray all-particle flux using a mass-composition assumption (proton or iron) and air shower simulations. If the fluxes differ, one can infer the data/MC ratio $R$ based on this equation:
  \begin{equation}
    J_\text{data}(\nmu_\text{data}) = J_\text{sim}(\nmu_\text{sim}) \, \frac{\text{d} \nmu_\text{sim}}{\text{d} \nmu_\text{data}} = J_\text{sim} \left( \frac{\nmu_\text{data}}{R} \right) \frac{1}{R}.
    \label{eq:ratio_from_flux}
  \end{equation}
  NEVOD-DECOR uses an average cosmic-ray flux computed from multiple experiments, while SUGAR uses the flux measured by the Pierre Auger Observatory. An energy is assigned to $R$ based on the muon density and $E \propto {\nmu}^{1/\beta}$.
\end{description}

\subsection{Energy scale offsets and cross-calibration}

If an experiment uses the same energy proxy for measurements of the cosmic-ray flux and the muon density, relative offsets in data/MC ratios can be corrected that arise purely from the different energy scales.

The cross-calibration uses that the cosmic-ray flux is very isotropic up to $10^{19.2}$\,eV and can serve as a universal reference. If we assume that all deviations in measured fluxes between different experiments arise from energy-scale offsets, then a relative energy-scale ratio $E_\text{data}/ E$ can be found for each experiment so that the all-particle fluxes overlap, based on an equation analog to \eq{ratio_from_flux}. This approach is well-known and has been used successfully in other works~\cite{Hoerandel:2002yg,Dembinski:2017zsh}.

\begin{table}
\centering
\caption{Energy-scale adjustment factors obtained from cross-calibration. The cross-calibration is explained in the text.}
\label{tab:energy_scale_correction}       
\begin{tabular}{lll}
\hline
Experiment & $E_\text{data} / E_\text{ref}$ \\
\hline
EAS-MSU & unknown \\
IceCube Neutrino Observatory & 1.19 \\
KASCADE-Grande & unknown \\
NEVOD-DECOR & 1.08 \\
Pierre Auger Observatory \& AMIGA & 0.948 \\
SUGAR & 0.948 \\
Telescope Array & 1.052 \\
Yakutsk EAS Array & 1.24 \\
\hline
\end{tabular}
\end{table}

\begin{figure}
\includegraphics[width=\columnwidth,trim=0 0 40 30]{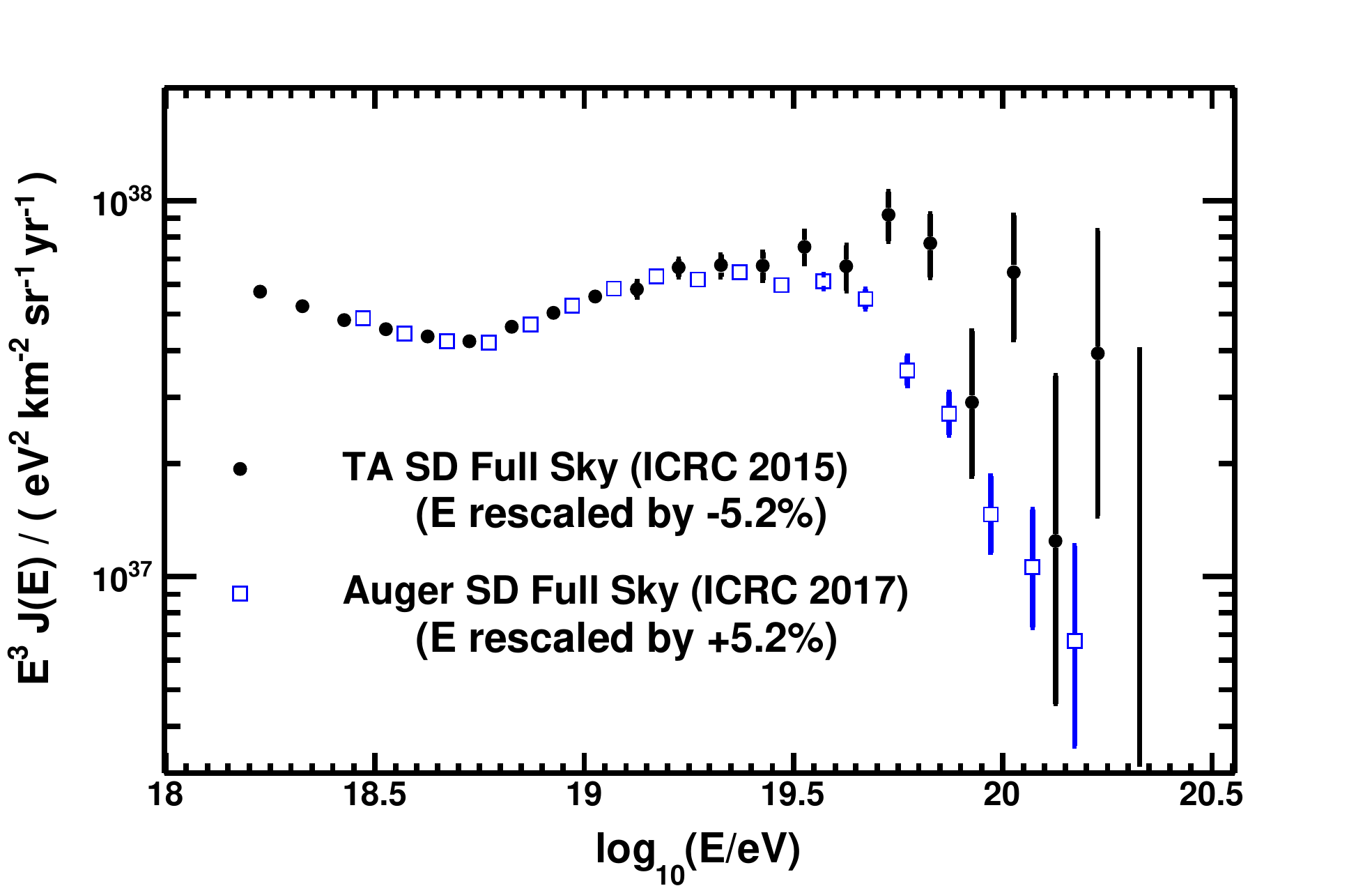}
\caption{Relative cross-calibration of the energy scales of the Pierre Auger Observatory and Telescope Array by matching flux measurements, as presented by the \emph{Spectrum Working Group}. Shown are the adjusted fluxes.}
\label{fig:spectrum_wg}
\end{figure}

\begin{figure}
\includegraphics[width=\columnwidth]{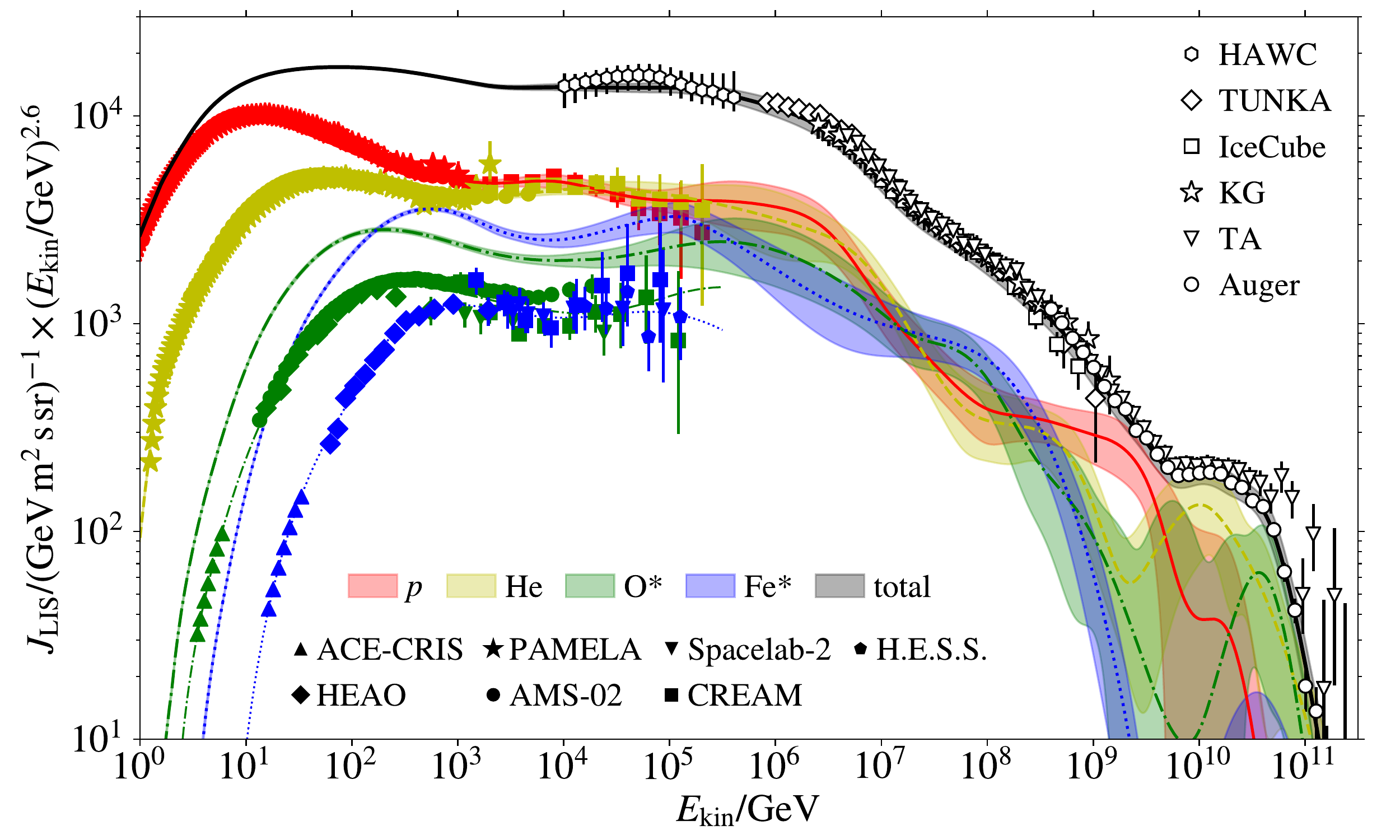}
\caption{Global spline fit (GSF) of cosmic-ray data (lines) and energy-scale adjusted data (points)~\cite{Dembinski:2017zsh}. The GSF treats energy-scale offsets as soft-constrained free parameters and produces energy-scale adjustment factors as a fit result.}
\label{fig:gsf}
\end{figure}

\begin{figure}
\includegraphics[width=\columnwidth,trim=30 20 30 35]{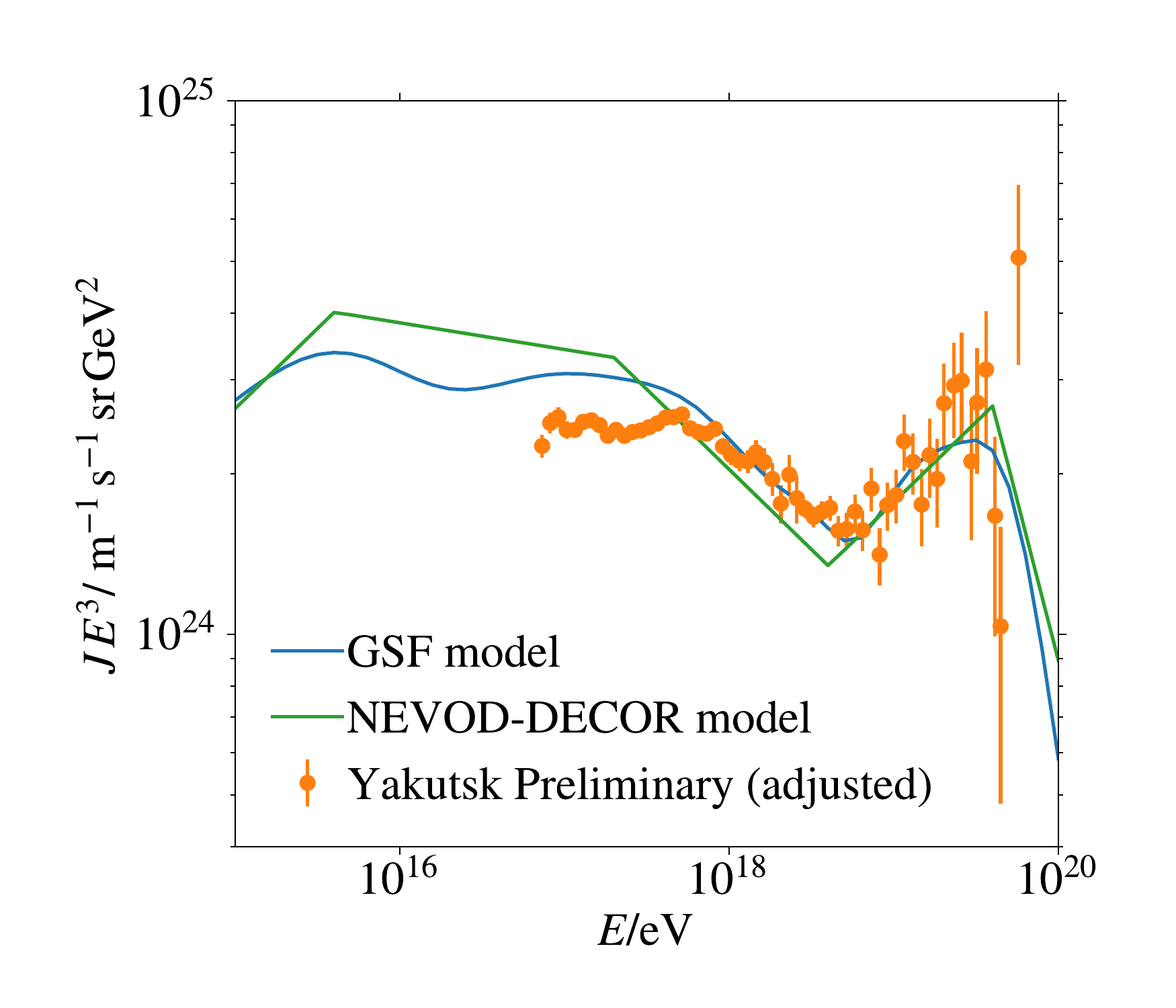}
\caption{All-particle flux from GSF~\cite{Dembinski:2017zsh} and NEVOD-DECOR~\cite{Bogdanov:2018sfw} models, and a preliminary update of the Yakutsk spectrum (see Pravdin et al. for the last publication~\cite{Pravdin:2009xah}), which was adjusted by an energy-scale factor 1.15 to match the GSF.}
\label{fig:yakutsk_nevod_decor_gsf}
\end{figure}

The cross-calibration factors used in this report are given in Table~\ref{tab:energy_scale_correction}. The \emph{Spectrum Working Group} formed by the Pierre Auger and Telescope Array collaborations found a relative energy-scale shift of 10.4\,\% between the two experiments, see \fg{spectrum_wg}. The reference energy scale $E_\text{ref}$ is placed between the two experiments.

The scaling factor for IceCube was taken from an updated Global Spline Fit (GSF) model~\cite{Dembinski:2017zsh}, see \fg{gsf}. The GSF also uses cross-calibration internally, with a reference energy scale $E_\text{ref,GSF}/E_\text{ref} = 0.948 / 0.88 \approx 1.08$. The scaling factor for IceCube is $E_\text{IceCube}/E_\text{ref,GSF} \times E_\text{ref,GSF}/E_\text{ref} = 1.10 \times 1.08 \approx 1.19$.

The factor for Yakutsk is obtained by matching a preliminary update of their all-particle flux against the GSF model, see \fg{yakutsk_nevod_decor_gsf}, to obtain a ratio $E_\text{Yakutsk}/E_\text{ref,GSF} \times E_\text{ref,GSF}/E_\text{ref} = 1.15 \times 1.08 \approx 1.25$. The factor for NEVOD-DECOR is obtained by comparing their custom all-particle flux parameterization with GSF (see same figure). The two models differ locally, which could be translated into energy-scale offsets within $\pm 2\,\%$, but no global offset is apparent. The energy scale of NEVOD-DECOR is therefore taken to be the same as GSF, $E_\text{NEVOD-DECOR}/E_\text{ref,GSF} \times E_\text{ref,GSF}/E_\text{ref} = 1 \times 1.08 = 1.08$.

No cross-calibration factor can be given for KASCADE-Grande, since the KASCADE-Grande flux is computed using a different energy estimator. For EAS-MSU, no all-particle flux is available for cross-calibration. SUGAR uses the flux from the Pierre Auger Observatory in its computation of the data/MC ratio and therefore has the same energy-scale adjustment factor.

We emphasize that the cross-calibration cannot eliminate a global offset of all experiments to the true energy scale, with corresponding shifts in the data/MC ratios. The energy scales of leading experiments have uncertainties in the order of 10 to 20\,\%, we assume that the reference energy-scale has an uncertainty of at least 10\,\%.

\subsection{Combined measurements}

\begin{figure*}
\centering
\includegraphics[width=\textwidth]{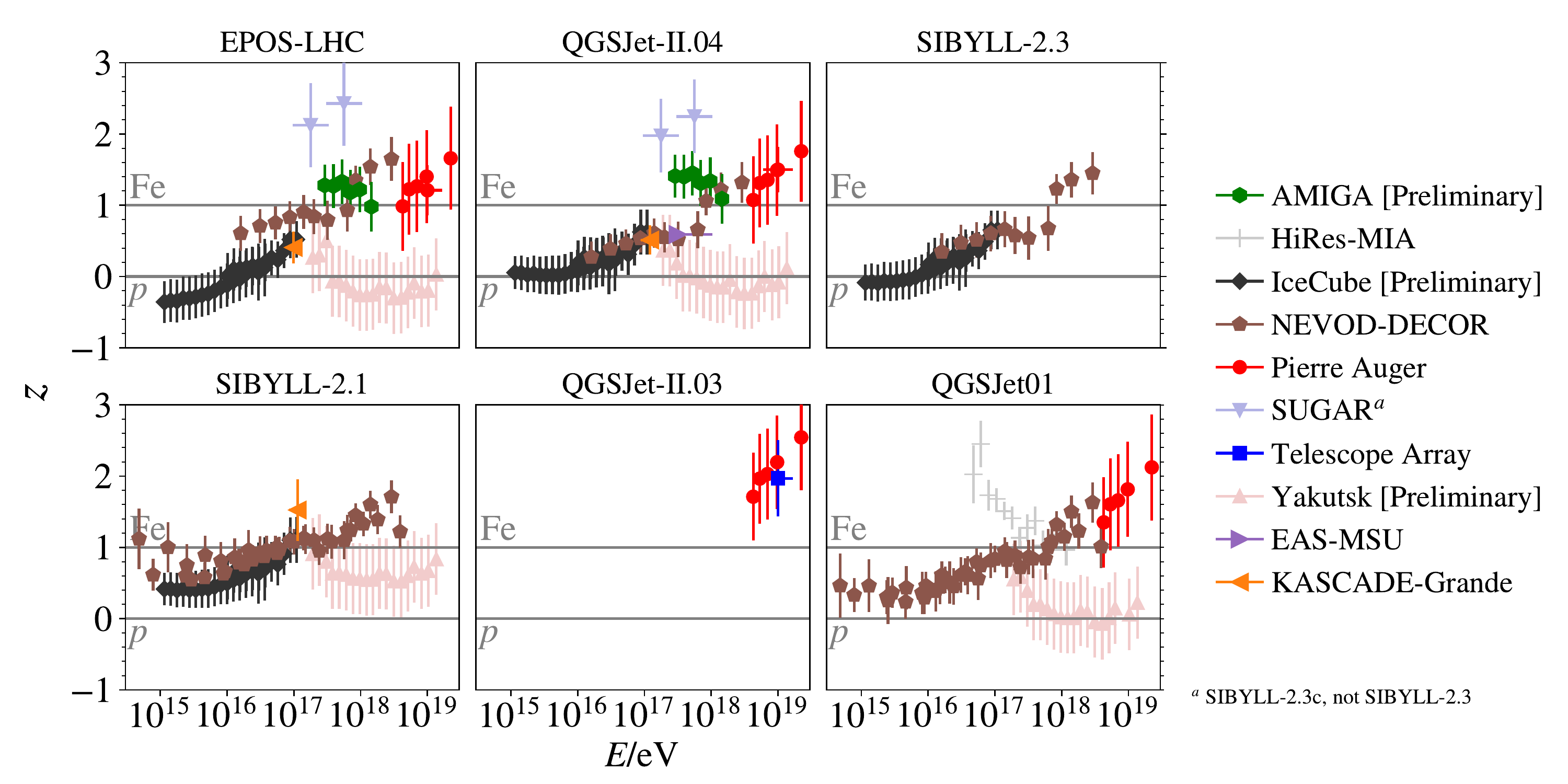}
\caption{Muon density measurements converted to the $z$-scale, as described in the text. It depends on the hadronic interaction model, so the same data sets are repeatedly shown for different models. When corresponding simulations are missing for an experiment, no points can be shown. Error bars show statistical and systematic uncertainties added in quadrature (systematic uncertainties are dominant for nearly all measurements).}
\label{fig:muon_original}
\end{figure*}

\begin{figure*}
\centering
\includegraphics[width=\textwidth]{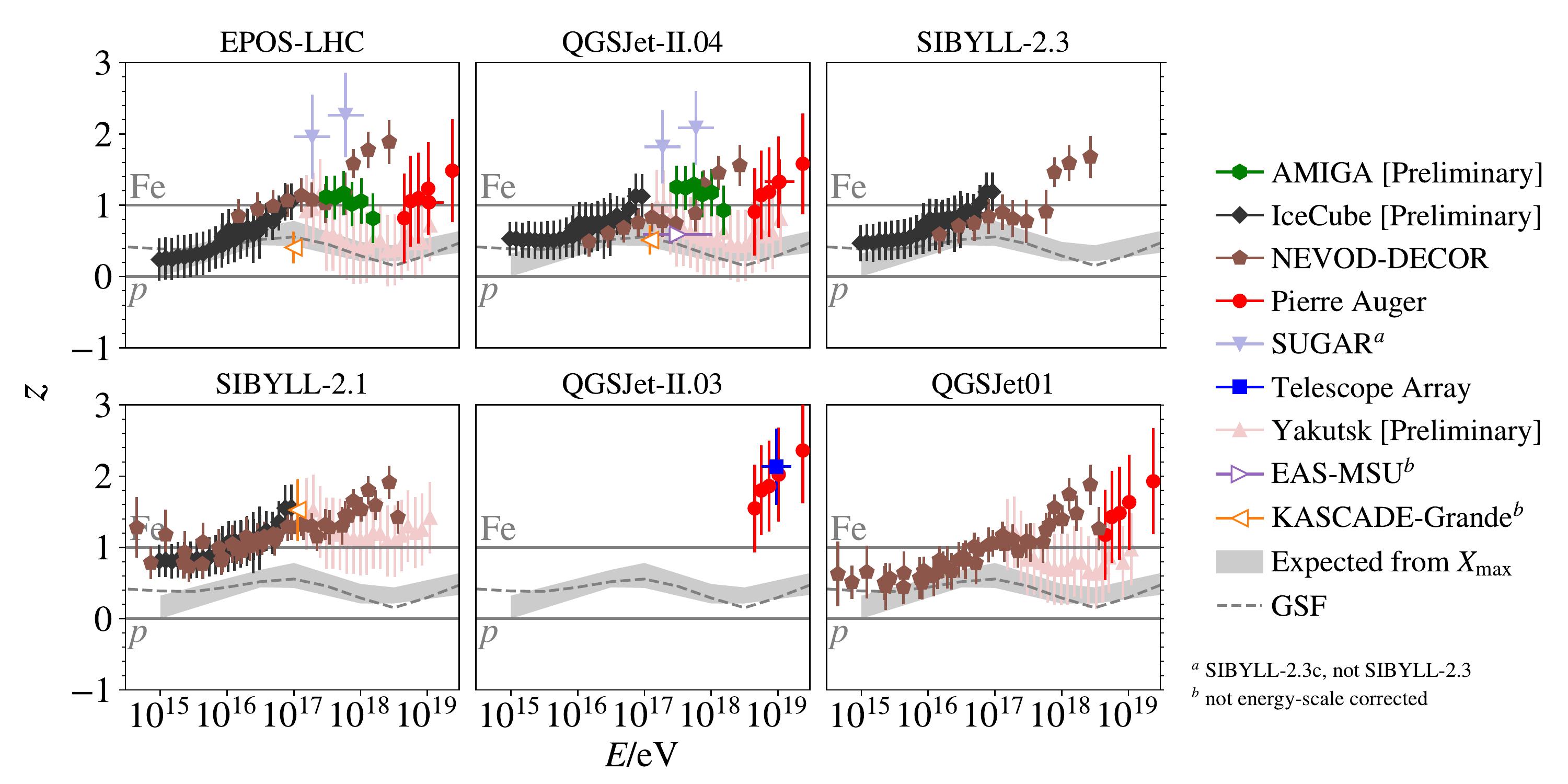}
\caption{Data from \fg{muon_original} after applying energy-scale cross-calibration. The points for KASCADE-Grande and EAS-MSU cannot be cross-calibrated and are only included for comparison. Shown for comparison are $z$-values expected for a mixed composition from optical measurements (band), based on an update of the review by Kampert and Unger~\cite{Kampert:2012mx} by the original authors of that paper, and from the GSF model (dashed line).}
\label{fig:muon_energy_rescaled}
\end{figure*}

\eq{lnnmu_e_a} displays a simple relationship between the measured muon density, $\mlna$ and logarithmic shower energy. To compare all the measurements, we introduce the $z$-scale, which is inspired by \eq{lnnmu_e_a},
\begin{equation}
z = \frac{\ln(\nmu^\text{det}) - \ln(\nmu_p^\text{det})}{\ln(\nmu_\text{Fe}^\text{det}) - \ln(\nmu_p^\text{det})},
\label{eq:z}
\end{equation}
where $\nmu^\text{det}$ is the muon density estimate as seen in the detector, while $\nmu_p^\text{det}$ and $\nmu_\text{Fe}^\text{det}$ are the simulated muon density estimates for proton and iron showers after full detector simulation. The $z$-scale, while being rather abstract, has advantages over other choices that were proposed. The energy-dependence of $\nmu$ is removed and the expected range is from 0 (pure proton showers) and 1 (pure iron showers), if there is no discrepancy between real and simulated air showers. This is convenient. Furthermore, biases of the form $\ln \nmu^\text{det} = A + B \ln \nmu$ in the measured muon density estimate $\nmu^\text{det}$ with respect to the true muon density $\nmu$ cancel in $z$.

Shown in \fg{muon_original} are the converted measurements. The $z$-values are computed relative to simulations and therefore a different result is obtained for each hadronic interaction model although the same data are used. The conversion to $z$ is only possible when $\nmu_\text{p}^\text{det}$ and $\nmu_\text{Fe}^\text{det}$ are available for that model. Therefore not all data points can be shown for all models. Overall, the data suggest an energy-dependent trend, but with a large scatter.

The scatter is drastically reduced after the cross-calibration, as shown in \fg{muon_energy_rescaled}. The cross-calibration causes a shift in the simulated values $\nmu_p$ and $\nmu_\text{Fe}$, which were computed for the energy $E_\text{data}$, but are needed for $E_\text{ref}$. Based on \eq{lnnmu_e_a}, we get $\ln \nmu_\text{ref} = \ln \nmu_\text{data} - \beta \ln (E_\text{data}/E_\text{ref})$. The shift is the same for proton and iron showers. It cancels in the denominator of \eq{z}, but enters with the opposite sign in the numerator. We get
\begin{equation}
  z_\text{ref} = z_\text{data} + \frac{\beta\ln (E_\text{data}/E_\text{ref})}{\ln(\nmu_\text{Fe}^\text{det}) - \ln(\nmu_p^\text{det})}
\end{equation}
with $\beta = 1 - (\ln\nmu_\text{Fe} - \ln\nmu_p)/\ln 56$, based on \eq{lnnmu_e_a}. The values of $\nmu_\text{Fe}$ and $\nmu_p$ are taken for each model from CORSIKA simulations. The points also move horizontally by the relative amount $(E_\text{data}/E_\text{ref})^{-1}$, a minor effect.

As expected, the cross-calibration improves the agreement of data from different experiments. Before and after the cross-calibration, the $z$-values rise above the iron line beyond $10^{19}$\,eV. The interpretation at lower energies changes, however. In case of IceCube, the originally negative $z$-values suggested that the muon density in proton showers simulated with EPOS-LHC for shower energies below $10^{16}$\,eV was too high. After the correction, the $z$-values fall between proton and iron. In case of Yakutsk, the original data suggested very low muon densities with partly negative $z$-values. After the correction, the Yakutsk data is consistent with others within uncertainties. We emphasize again that the reference energy-scale after cross-calibration has a remaining uncertainty of at least 10\,\%. This means that $z$-values in all plots can be collectively varied by about $\pm 0.25$. 

To further refine the conclusions, we consider the effect of an energy-dependent mass composition. With \eq{lnnmu_e_a} and \eq{z} the expected value $z_\text{mass}$ for a given mean-logarithmic-mass $\mlna$ is computed as $z_\text{mass} = \frac{\mlna}{\ln 56}$.
As mentioned in the introduction, the experimental value of $\mlna$ is uncertain. Shown in \fg{muon_energy_rescaled} is a band, an envelope over optical measurements of the depth $\xmax$ of shower maximum from several experiments, and converted to $\mlna$ based on air shower simulations with EPOS-LHC. We will use this as a rough estimate of the mass composition. The band is independent of the muon measurements here, and therefore can be used as a reference. The $z_\text{mass}$ value computed from the GSF model is also shown, which is based on optical and muon measurements and averages over experiments and model interpretations of air shower data. The line mostly falls inside the envelope.

\begin{figure*}
\centering
\includegraphics[width=0.9\textwidth]{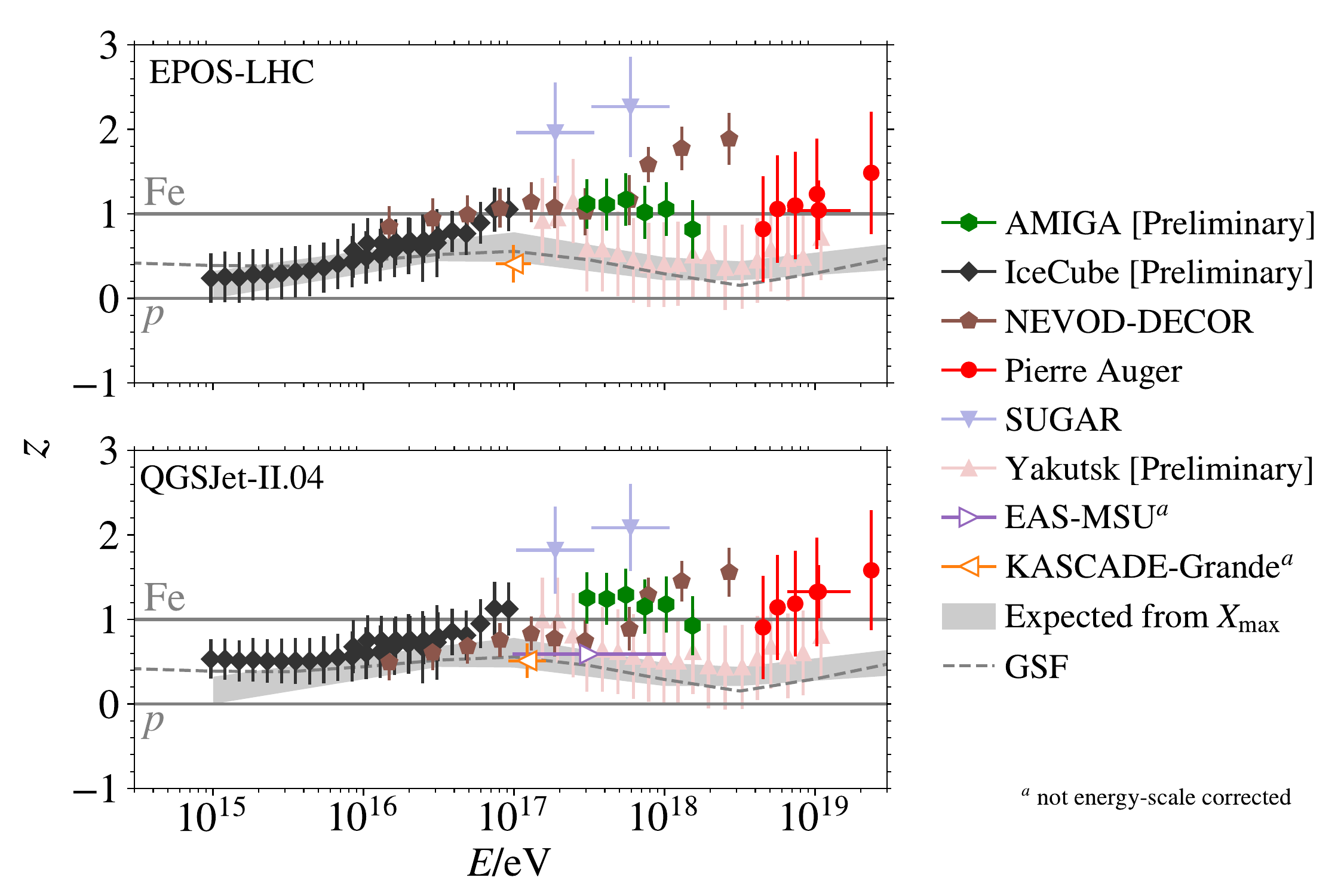}
\caption{Zoom into \fg{muon_energy_rescaled} for EPOS-LHC and QSGJet-II.04. The points for KASCADE-Grande and EAS-MSU cannot be cross-calibrated and are only included for comparison.}
\label{fig:zoomed}
\end{figure*}

\begin{figure*}
\centering
\includegraphics[width=0.9\textwidth]{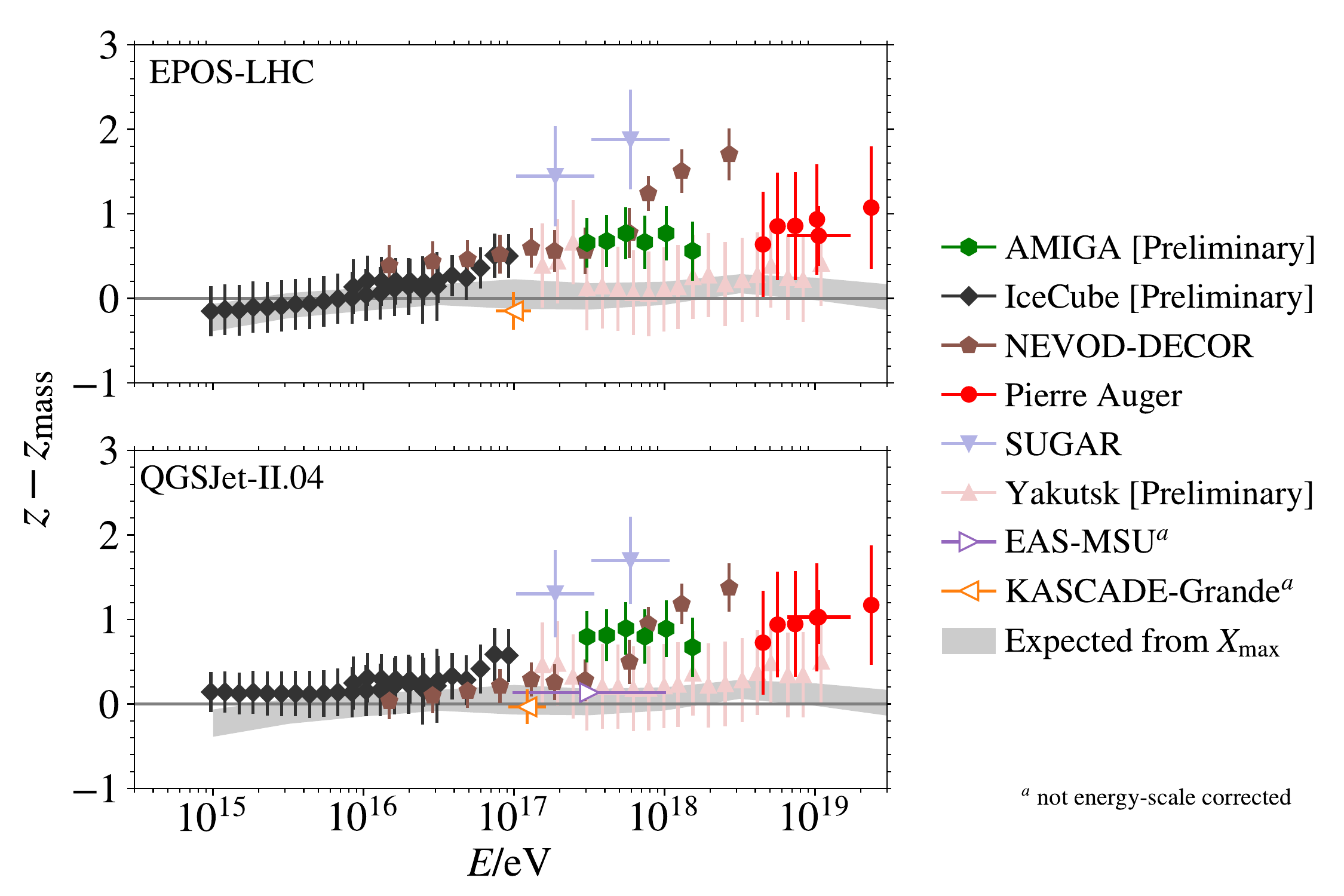}
\caption{Data from \fg{zoomed} after subtracting $z_\text{mass}$.  The points for KASCADE-Grande and EAS-MSU cannot be cross-calibrated and are only included for comparison.}
\label{fig:zoomed_subtracted}
\end{figure*}

If the measured $z$ values follow $z_\text{mass}$, the model describes the muon density at the ground consistently. This is overall not the case. The pre-LHC generation of hadronic interaction models, SIBYLL-2.1, QGSJet-II.03, and QGSJet01~\cite{Kalmykov:1993qe}, show larger muon deficits than the models tuned to LHC data, EPOS-LHC, QGSJet-II.04, and SIBYLL-2.3. EPOS-LHC, QGSJet-II.04, SIBYLL-2.3, and QGSJet01 give a reasonable description of data up to a few $10^{16}$\,eV. At higher shower energies, a muon deficit in simulations is observed ($z > z_\text{mass}$) in all models. Shown in \fg{zoomed} are zoomed plots for EPOS-LHC and QGSJet-II.04, the two latest-generation models with most data points. Shown in \fg{zoomed_subtracted} is the difference $\Delta z = z - z_\text{mass}$. Subtracting $z_\text{mass}$ is expected to remove the effect of the changing mass composition. An energy-dependent trend in $\Delta z$ remains.

\subsection{Energy-dependent trend}

\begin{figure*}
\centering
\includegraphics[width=0.49\textwidth,clip,trim=30 10 30 10]{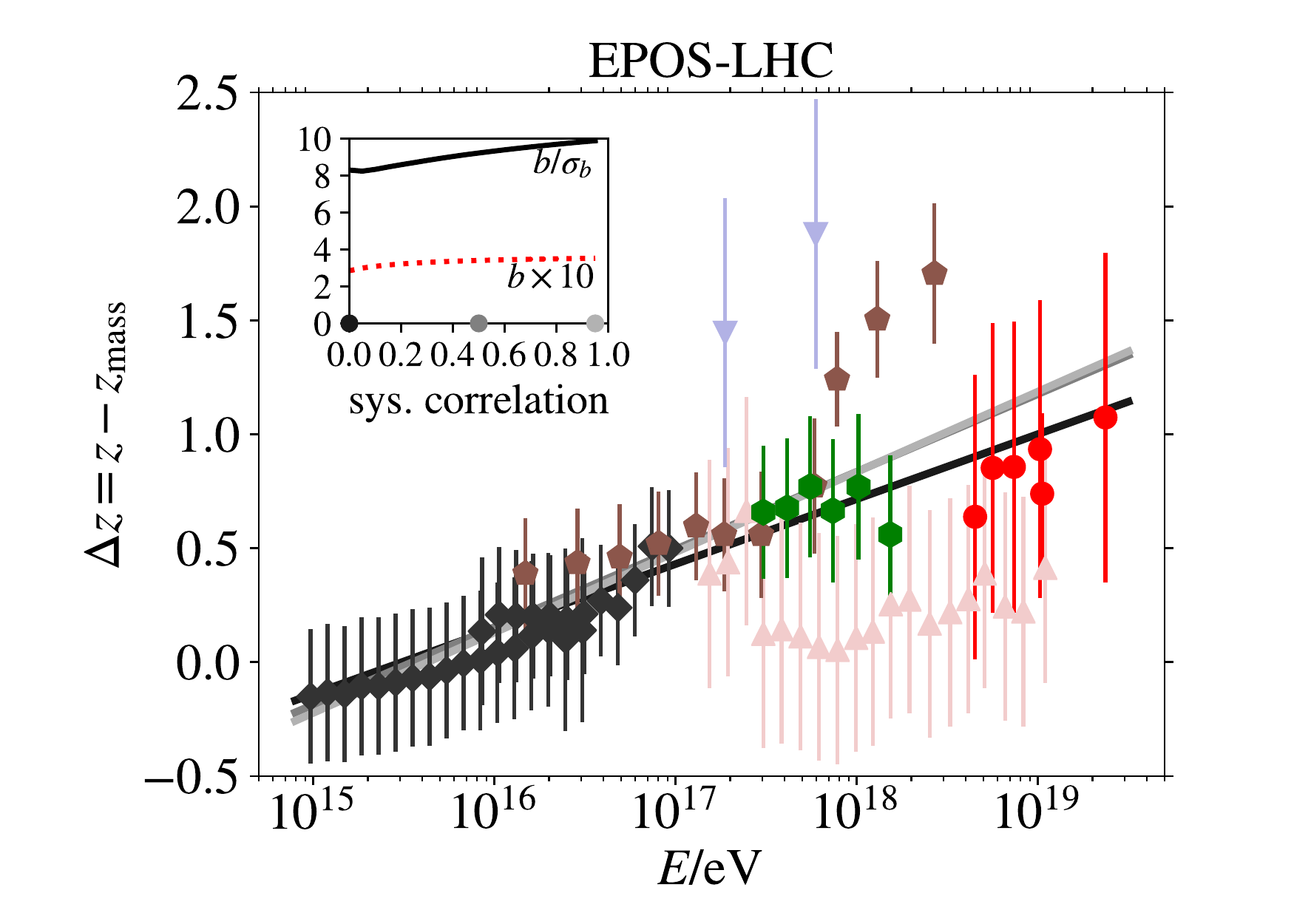}
\hfill
\includegraphics[width=0.49\textwidth,clip,trim=30 10 30 10]{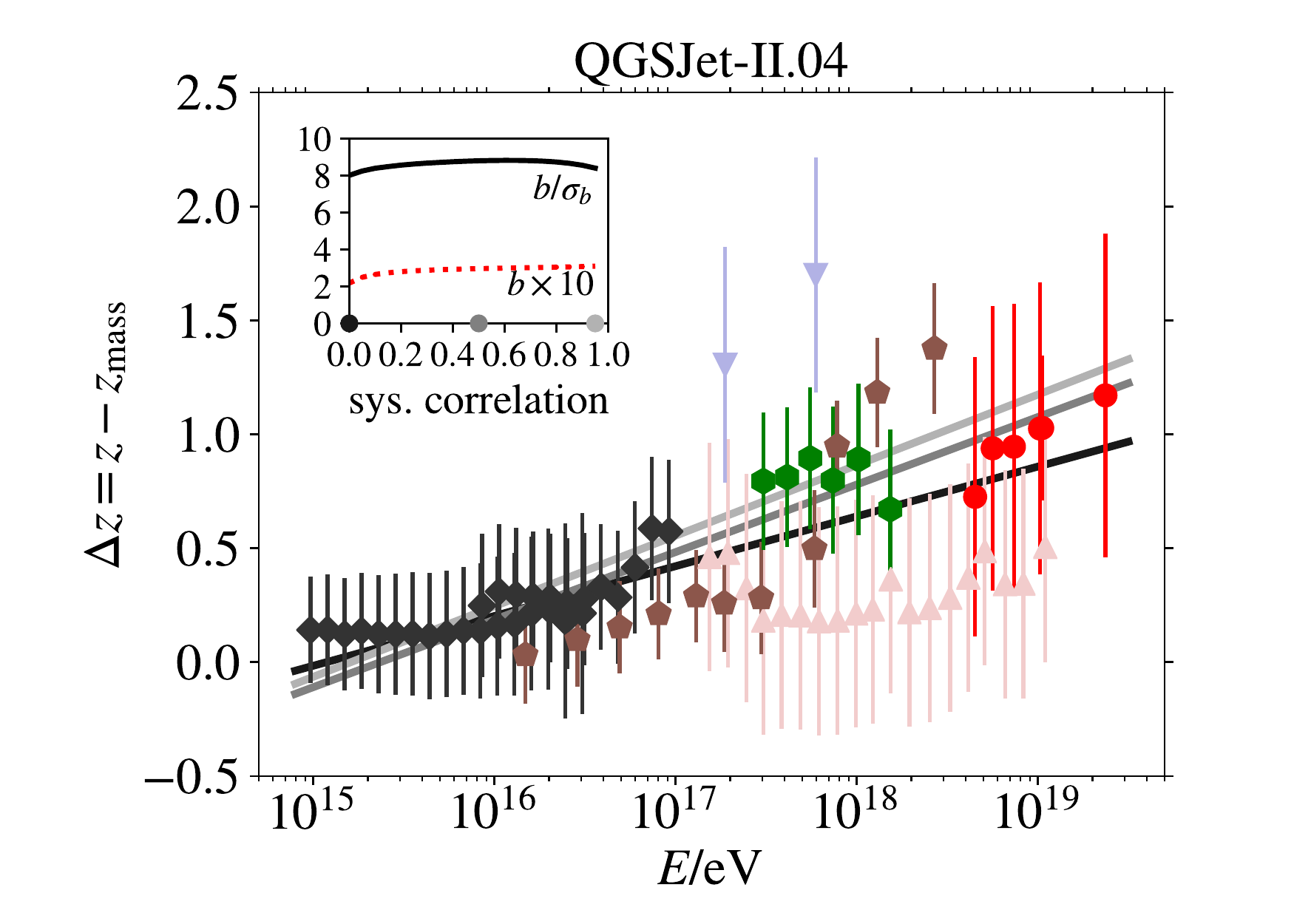}
\caption{Fits of straight lines $\Delta z = a + b \, (\log_{10}(E/\text{eV}) - 16)$ to the data points. Shown in the inset are the slope $b$ and its deviation from zero in standard deviations for an assumed correlation of the point-wise uncertainties within each experiment. Examples of the fitted lines are shown for a correlation of 0, 0.5, and 0.95 in varying shades of gray.}
\label{fig:zfit}
\end{figure*}

To quantify the observed trend in $\Delta z$ as a function of energy, a line-model is fitted to the data shown in \fg{zoomed_subtracted},
\begin{equation}
\Delta z = a + b \, (\log_{10}(E/\text{eV}) - 16),
\label{eq:fit}
\end{equation}
with free parameters $a$ and $b$. The slope $b$ is the increase in $\Delta z$ per decade in energy. The $z$-values from KASCADE-Grande and EAS-MSU are not included in the fit, since they are not energy-scale corrected and not directly comparable to the other values.

The value of the slope $b$ and its deviation from zero in standard deviations are of interest. The uncertainty of $b$ scales with the uncertainties of the data. The error bars of most data points are dominated by systematic uncertainties, which are correlated for data points from a single set. Correlated uncertainties are the reason why the points in \fg{zoomed_subtracted} do not scatter randomly as much as the error bars suggest. The exact amount of correlation is not known. We work around this problem by repeatedly fitting the data under different correlation assumptions.

We use the least-squares method for correlated uncertainties in the data. The score $Q$ is minimized,
\begin{equation}
Q = (\boldsymbol{\Delta z} - \boldsymbol{\Delta z}^\text{line}) ^T \, C^{-1} \, (\boldsymbol{\Delta z} - \boldsymbol{\Delta z}^\text{line}),
\end{equation}
where $C^{-1}$ is the inverse of the covariance matrix $C$ of the measurements. The matrix $C$ is constructed as follows
\begin{equation}
  C_{ij} = \sigma_{\Delta z,i} \, \sigma_{\Delta z,j} \times
  \begin{cases}
    1 & \text{for } i = j \\
    \alpha & \text{for }i,j \text{ from same data set} \\
    0 & \text{otherwise},
  \end{cases}
\end{equation}
where $\alpha$ is the assumed correlation coefficient for the error bars within a data set. The matrix $C$ has a block-diagonal form.

The fit is performed with MINUIT~\cite{James:1975dr} and repeated for values of $\alpha$ from 0 to 0.95. The HESSE algorithm is used to compute the standard deviation $\sigma_b$ of $b$. To adjust for over- or underestimated uncertainties in the input, the raw result from MINUIT is corrected with the $\chi^2$ value and the degrees of freedom $n_\text{dof}$ of the fit, $\sigma_b = \sigma_b^\text{raw} \sqrt{\chi^2/n_\text{dof}}$~\cite{James:2006zz}.

The slope $b$ and its deviation from zero in standard deviations as a function of the assumed correlation are shown in \fg{zfit} for EPOS-LHC and QGSJet-II.04. The deviation from zero is always larger than 8 standard deviations, making the slope highly significant. The result is insensitive to the assumed correlation coefficient.

\section{Summary and outlook}

We presented a summary of recent tests and measurements of hadronic interaction properties in air showers with energies from PeV up to tens of EeV. Better agreement between simulation and experiment is found for the electromagnetic than for the muonic shower component.

We put a special focus on muon density measurements in this report. A comprehensive collection of muon measurements is presented. We developed the $z$-scale as a comparable measure of the muon density between different experiments and analyses. The $z$-scale uses air shower simulations as a reference to compare muon density measurements taken under different conditions. We demonstrated the importance of cross-calibrating energy-scales of experiments and apply it were possible, using the isotropic all-particle flux of cosmic rays as a reference.

After applying the cross-calibration, a remarkably consistent picture is obtained. Muon measurements seem to be consistent with simulations based on the latest generation of hadronic interaction models, EPOS-LHC, QGSJet-II.04, and SIBYLL-2.3, up to about $10^{16}$\,eV. At higher energies, a growing muon deficit in the simulations is observed, visible as an increase in $z$ over the expectation. An analog trend is observed in older hadronic interaction models, with a more severe muon deficit. The slope of this increase in $z$ per decade in energy is 0.22 to 0.35 for EPOS-LHC and QGSJet-II.04, with 8\,$\sigma$ significance.

We plan to further study the collected data, looking for other trends in the deviation between simulations and data. This will provide hints which aspects of hadronic interaction properties are the likely cause for these deviations.

\section{Acknowledgements}

We remember Mikhail Pravdin of the Yakutsk collaboration, who passed away on January 3rd, 2019. Dr.~Pravdin had a leading role in cosmic ray physics. His kindness, skills, and expertise will be greatly missed.


\bibliography{refs}

\end{document}